\documentclass[utf8]{frontiersFPHY} 

\setcitestyle{square} 
\usepackage{url,hyperref,lineno,microtype,subcaption}
\usepackage[onehalfspacing]{setspace}

\usepackage[utf8]{inputenc}
\usepackage{graphicx}
\usepackage{lipsum}
\usepackage{lettrine}
\usepackage{pdflscape}
\usepackage{amsmath}
\usepackage{amssymb}
\usepackage{mathtools}
\usepackage{hyperref}
\usepackage{aasmacros}
\usepackage{array}
\usepackage{float}
\usepackage{ragged2e}
\usepackage{ulem}
\usepackage{color}
\usepackage{capt-of}
\usepackage{mathtools, cuted}

\newcolumntype{L}[1]{>{\raggedright\let\newline\\\arraybackslash\hspace{0pt}}m{#1}}
\newcolumntype{C}[1]{>{\centering\let\newline\\\arraybackslash\hspace{0pt}}m{#1}}
\newcolumntype{R}[1]{>{\raggedleft\let\newline\\\arraybackslash\hspace{0pt}}m{#1}}
\newcolumntype{P}[1]{>{\centering\arraybackslash}p{#1}}
\makeatletter
\def\cl@chapter{\@elt {theorem}}
\makeatother
\graphicspath{{Figures/}}

\newcommand{\norm}[1]{\left\lVert#1\right\rVert}

\DeclareUnicodeCharacter{00A0}{ }

\renewcommand{\emph}[1]{{\it #1}}

\def\keyFont{\fontsize{8}{11}\helveticabold }
\def\firstAuthorLast{Perraudin {et~al.}} 
\def\Authors{
Nathana\"el Perraudin\,$^{1}$, 
Sandro Marcon\,$^{2}$,
Aurelien Lucchi\,$^{2}$,
Tomasz~Kacprzak\,$^{3,*}$
}

\def\Address{
$^{1}$Swiss Data Science Center, ETH Zurich \\
$^{2}$Institute for Machine Learning, ETH Zurich, \\ 
$^{3}$Institute for Particle Physics and Astrophysics, ETH Zurich }
\def\corrAuthor{Tomasz Kacprzak}
\def\corrEmail{tomaszk@phys.ethz.ch}

\begin{document}
\onecolumn
\firstpage{1}

\title[Cosmological mass map emulation with CGANs]{Emulation of cosmological mass maps with conditional generative adversarial networks} 

\author[\firstAuthorLast ]{\Authors} 
\address{} 
\correspondance{} 

\extraAuth{}

\maketitle

\begin{abstract}

\noindent
Weak gravitational lensing mass maps play a crucial role in understanding the evolution of structures in the universe and our ability to constrain cosmological models.
The prediction of these mass maps is based on expensive N-body simulations, which can create a computational bottleneck for cosmological analyses. 
Simulation-based emulators of map summary statistics, such as the matter power spectrum and its covariance, are starting to play increasingly important role, as the analytical predictions are expected to reach their precision limits for upcoming experiments.
Creating an emulator of the cosmological mass maps themselves, rather than their summary statistics, is a more challenging task.
Modern deep generative models, such as Generative Adversarial Networks (GAN), have demonstrated their potential to achieve this goal.
Most existing GAN approaches produce simulations for a fixed value of the cosmological parameters, which limits their practical applicability. 
We propose a novel conditional GAN model that is able to generate mass maps for any pair of matter density~$\Omega_m$ and matter clustering strength~$\sigma_8$, parameters which have the largest impact on the evolution of structures in the universe, for a given source galaxy redshift distribution $n(z)$. 
Our results show that our conditional GAN can interpolate efficiently within the space of simulated cosmologies, and generate maps anywhere inside this space with good visual quality high statistical accuracy.
We perform an extensive quantitative comparison of the N-body and GAN -generated maps using a range of metrics:
the pixel histograms, peak counts, power spectra, bispectra, Minkowski functionals, correlation matrices of the power spectra, the Multi-Scale Structural Similarity Index (MS-SSIM) and our equivalent of the Fr\'echet Inception Distance (FID).
We find a very good agreement on these metrics, with typical differences are $<$5\% at the centre of the simulation grid, and slightly worse for cosmologies at the grid edges. 
The agreement for the bispectrum is slightly worse, on the $<$20\% level.
This contribution is a step towards building emulators of mass maps directly, capturing both the cosmological signal and its variability.
We make the code\footnote{\url{https://renkulab.io/gitlab/nathanael.perraudin/darkmattergan}} and the data\footnote{\url{https://zenodo.org/record/4646764}} publicly available.
\vspace{0.5em}
\tiny
\keyFont{ \section{Keywords:} 
generative models,
cosmological simulations,
cosmological emulators,
N-body simulations,
mass maps,
generative adversarial network,
fast cosmic web simulations,
conditional GAN
} 
\end{abstract}

\newpage
\section{Introduction}

The N-body technique simulates the evolution of the universe from soon after the big bang, where the mass distribution was approximately a Gaussian random field, to today, where, under the action of gravity, it becomes highly non-Gaussian. 
The result of an N-body simulation consists of a 3D volume where the positions of particles represent the density of matter in specific regions.
This 3-dimensional representation can then be projected in 2 dimensions by integrating the mass along the line of sight with a lensing kernel. The resulting images are called \emph{sky convergence maps}, often referred to simply as the \emph{cosmological mass maps}. 
These maps can be compared with real observations with the purpose of estimating the cosmological parameters and testing cosmological models.
Their simulation, however, is a very challenging task: a single large N-body simulation can take from a few hours to several weeks on a supercomputer \cite{ euclid2019euclid,potter2017pkdgrav3,springel2005millenium, sgier2019fast}.

One approach to overcome this challenge is to use simulation-based emulators of summary statistics of the maps. Emulators have so far focused on: (a) the power spectrum, which is commonly used in cosmology~\citep{knabenhans2019euclidemu,Heitmann2016cosmicemu,Knabenhans2020euclidemulator2,Angulo2020Bacco}, (b) covariance matrices of 2-pt functions \cite{Taylor2013precision,Sato2011covariance,Sgier2019covariance}, and (c) non-Gaussian statistics of mass maps, which can be a source of significant additional cosmological information~\cite{Pires2009discrimination,Petri2013minkowski,Zuercher2020forecast,fluri2018cosmological}.
These approaches, however, always considered a specific summary statistic, which limits the type of analysis that can be performed using the mass-map data.
They typically do not simultaneously capture both the signal and its variation: the emulators interpolate the power spectrum across the cosmological parameter space, without considering the change in its covariance matrix, which is typically taken from the fiducial cosmology parameter set.
This is a known source of potential error in the analysis~\cite{Eifler2009covariance} and was shown to have a large impact on the deep learning-based constraints~\cite{fluri2018cosmological}.
The solution proposed in this work address these problems simultaneously. 
We construct a map-level probabilistic emulator that generates the mass maps directly, and can accurately capture the signal and its variability.
This emulator, built for a specific target survey dataset, would be of great practical use for innovative map-based cosmological analyses, additionally capturing the variation of the maps across the cosmological parameter space.

With a similar goal, multiple contributions have leveraged the recent advances in the field of deep learning to aid the generation of cosmological simulations.
In particular, recent works \cite{mustafa2017creating,rodriguez2018fast,perraudin2019cosmological,Troester2019painting} have demonstrated the potential of Generative Adversarial Networks (GAN) \cite{goodfellow2014generative} for production of N-body simulations.
The work of \cite{mustafa2017creating,rodriguez2018fast,perraudin2019cosmological,Troester2019painting,Guisarma2019neutrino,He2019learning} has shown deep generative models that can accurately model dark matter distributions and other related cosmological signals.
However, a practical application of these approaches in an end-to-end cosmological analysis is yet to be demonstrated.
In this work, we take an essential step towards the practical use of generative models by creating the first emulator of weak lensing mass maps as function of cosmological parameters. This step allows the generate mass maps with any parameters without the need to retrain the generative model.
Our conditional GAN model generates convergence maps dependent on values of two parameters that have the largest impact on the evolution of the Large Scale Structure (LSS) of the universe: $\Omega_m$, which controls the matter density as a fraction of total density, and $\sigma_8$, which controls the strength of matter density fluctuations (see \citep{Refregier2003review,Kilbinger2015review} for reviews).
Those are the only two parameters that can be effectively measured using the convergence maps data.
After training, the conditional model can then interpolate to unseen values of $\sigma_8$ and $\Omega_m$ by varying the distribution of the input latent variable. 
Other works \cite{Tamosiunas2020investigating,VillaescuseNavarro2020camels} have since also explored such models, although with the emphasis on generating various cosmological fields themselves, either in 2D or 3D.

To assess that the GAN-generated maps are statistically very close to the originals, we perform an extensive quantitative comparison. 
We evaluate our GAN using both cosmological and image processing metrics: 
the power spectral density,
mass map histogram, 
peak histogram, 
the bispectrum, 
Minkowski functionals,  
Multi-Scale Structural Similarity (MS-SSIM)~\cite{wang2003multiscale}, 
and an adaptation of the Fr\'echet Inception Distance (FID) \cite{heusel2017gans}.
We also compare the statistical consistency of a batch of generated maps by computing the correlation matrices of power spectra.
Moreover, we assess the agreement as a function of cosmological parameters.
This set of comparisons is the most exhaustive presentation of the capacity of generative models to learn the dark matter maps, to date.
In this work we use the data generated by~\cite{fluri2019kids}.

We build a sky convergence map dataset made of $57$ different cosmologies (set of parameters) divided into a training set and a test set. 
The test set consists of 11 cosmological parameters sets was used to asses the capacity of the GAN to interpolate to unseen cosmologies.

This paper is structured as follows.
In Section~\ref{sec:conditional_gans} we present a new type of generative adversarial network whose generated output can be conditioned on a set of parameters in the form of continuous values.
Section~\ref{sec:data_kids} describes the simulation dataset used in this work.
In Section~\ref{sec:metrics} we describe the metrics used to evaluate the quality of the generative model.
Section~\ref{sec:kids_conditional_results} shows the maps generated by our machine learning model, as well as compares its results to the original, simulated data.
We summarise our findings and discuss the future prospects in Section~\ref{sec:conclusion}.
Appendix A contains the architectures of the neural networks used in this work.

\section{Conditional generative adversarial networks}
\label{sec:conditional_gans}
A GAN consists of two neural networks, $D$ and $G$, competing against each other in a zero-sum game. 
The task of the \emph{discriminator} $D$ is to distinguish real (training) data from fake (generated) data. 
Meanwhile, the \emph{generator} $G$ produces samples with the goal of deceiving the discriminator into believing that the generated data is real. 
Both networks are trained simultaneously and if the optimization process is carried out successfully, the generator will learn to produce the data distribution \cite{goodfellow2014generative}. 
Learning the optimal parameters of the discriminator and generator networks can be formulated as optimizing a min-max objective. 
Optimizing a GAN is a challenging task due to the fact that it consists of two networks competing against each other. 
In practice, one often observes unstable training behaviours which can be mitigated by relying on various types of regularization methods~\cite{roth2017stabilizing, gulrajani2017improved}. 
In this paper, we rely on Wasserstein GANs~\cite{arjovsky2017wasserstein} with the regularization approach suggested in~\cite{gulrajani2017improved}. 
The model we use conditions both the generator and the discriminator on a given random variable $y$, yielding the following objective function,
\begin{equation}
\label{eq:gulrajani_Loss}
    \min_G \max_D \mathop{\mathbb{E}}_{(\pmb{x},\pmb{y}) \sim \mathbb{P}_r}[D(\pmb{x},\pmb{y})] - \mathop{\mathbb{E}}_{\pmb{z} \sim \mathbb{P}_z, \pmb{y} \sim \mathbb{P}_y}[ D(G(\pmb{z}, \pmb{y}))]
    + \lambda \mathop{\mathbb{E}}_{(\pmb{x},\pmb{y}) \sim \mathbb{P}_r \cup \mathbb{P}_g}[( \| \nabla_{\pmb{x}} D(\pmb{x}, \pmb{y} )\|_{2} - 1)^2,
\end{equation}
where $\mathbb{P}_r$ and $\mathbb{P}_z$ are the data and latent variable distributions.  
The parameter $\lambda \geq 0$ is the penalty coefficient of the regularization term that ensures that the gradient norm of the discriminator is close to $1$. 
This ensures that the discriminator is 1-Lipschitz, which is a requirement for optimizing the Wasserstein distance~\cite{arjovsky2017wasserstein,gulrajani2017improved}.
The prior distribution of the latent variable, e.g., a uniform or a Gaussian distribution, defines implicitly the generator distribution $\mathbb{P}_g$ by $(\pmb{x}, \pmb{y}) = G(\pmb{z},\pmb{y}),\pmb{z} \sim \mathbb{P}_z,\pmb{y} \sim \mathbb{P}_y $.

\begin{figure*}[t!]
\centering
\includegraphics[width=0.9\textwidth]{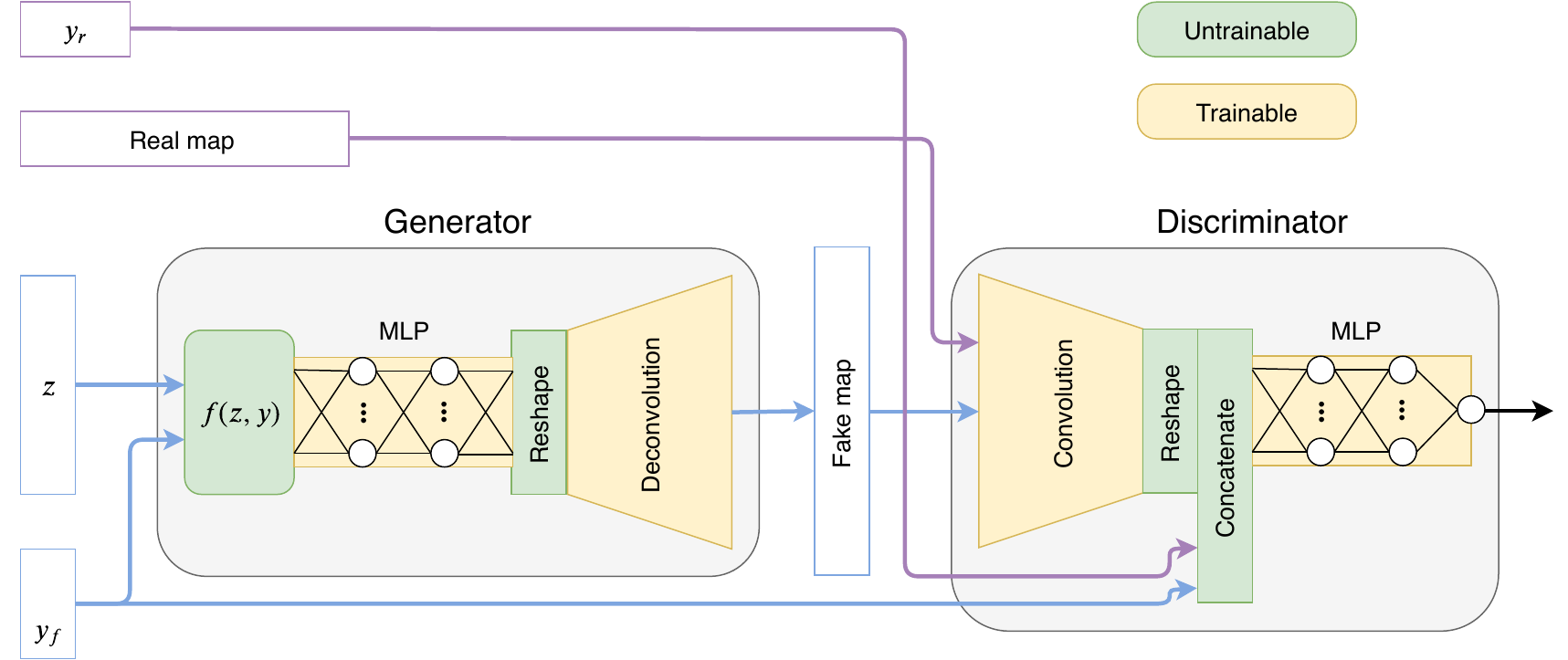}
\caption{Sketch of the proposed GAN model, where $z$ is a latent variable and $y$ is a parameter vector ($y_r$~=~real, $y_f$~=~fake).}
\label{fig:our_model}
\end{figure*}

Practically, there exist many techniques and architectures to condition the generator and the discriminator \cite{gauthier2014conditional, reed2016generative, perarnau2016invertible,odena2017conditional, miyato2018cgans}. 
However, all the architectures in these works are conditioning on discrete parameters.
We instead propose a different design that works specifically for continuous parameters and will be shown to have good performance in practice. We note that our conditioning technique could be used with other architectures as well.
For simplicity we describe the case of a single parameter, but our technique was implemented for the case of two parameters.
Our idea is to adapt the distribution of the latent vector according to the conditioning parameters using the function $\hat{z} = f(z, y)$. Specifically, the function $f$ simply rescales the norm of the latent vector according to the parameter $y$. Given the range $y \in [a, b]$, $f$ reads:
\begin{equation}
\hat{z} = f(z, y) = \left( l_0 + \frac{l_1 - l_0}{b - a} (y - a) \right) \frac{z}{\|z\|_2}.
\label{eq:gaussian_length}
\end{equation}
Using this function, the length of the $z$ vector is mapped to the interval $[l_0, l_1]$. In our case, we used $l_0 = 0.1 \sqrt{n}$ and $l_1 = \sqrt{n}$, where $n$ is the size of the latent vector. 
For the discriminator, the parameters are concatenated directly after the convolutional layers as in~\cite{reed2016generative}.
The relation between the features extracted from the convolutional layers and the parameters might in general be non-local. 
We therefore increase the complexity of the mapping functions of the discriminator and generator by adding some linear layers (as in a multi-layer perceptron) at the end of the discriminator and the beginning of the generator. 
The proposed model is sketched in Figure~\ref{fig:our_model} and the architecture is described in more details in Appendix A. Specific parameters can be found in Table \ref{tab:kids_gan_conditional}.

\section{Sky convergence maps dataset}
\label{sec:data_kids}

\begin{figure}
  \centering
%
  \includegraphics[clip, trim=0cm 0cm 0cm 0cm, width=0.4\linewidth]{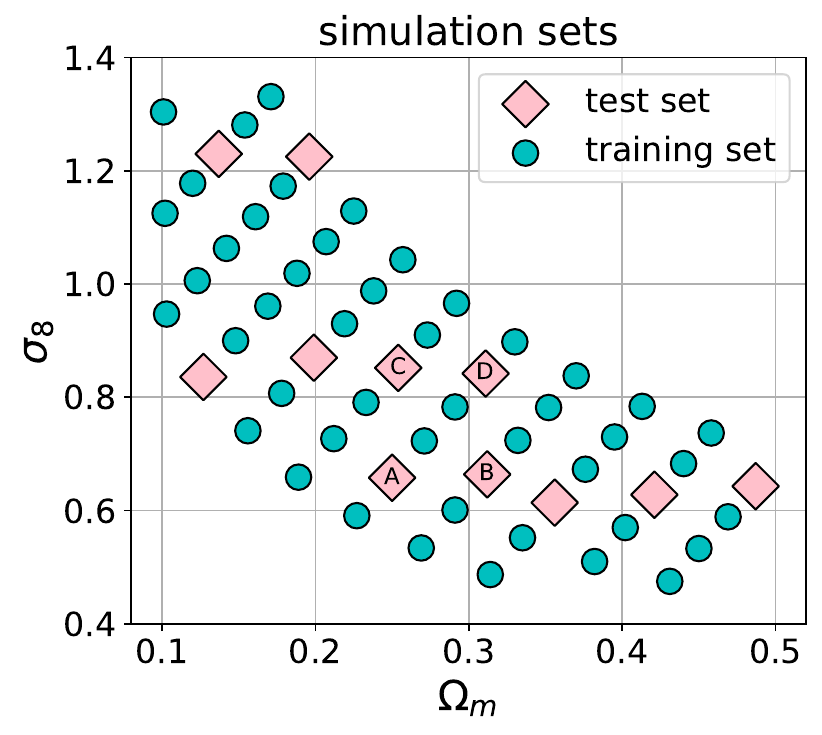}
  \caption{
  The cosmological parameter grid used in this work, from \citep{fluri2019kids}.
  The circles and diamonds show the training and the test sets, respectively.
  The total number of models was 57, of which 46 were used as the training set and 11 as the test set.
  The models labelled A,B,C, and D are investigated in more detail in Section~\ref{sec:kids_conditional_results}.
  }
  \label{fig:grid_test_train}
\end{figure}

The data used in this work is the non-tomographic training and testing set introduced in~\citep{fluri2019kids}, without noise and intrinsic alignments.
The simulation grid consists of $57$ different cosmologies in the standard cosmological model: a flat universe with cold dark matter ($\Lambda$CDM) \cite{LahavLiddle2019parameters}. 
Each of these $57$ configurations was run with different values of $\Omega_m$ and $\sigma_8$, resulting in the parameter grid shown in Figure \ref{fig:grid_test_train}.\
The output of the simulator consists of the particle positions in 3D space. 
The mass maps are obtained by the \emph{gravitational lensing} technique (see \cite{Bartelmann2010review} for review).
It consists of a tomographic projection of the particle densities along the radial  (redshift)  direction against the \emph{lensing kernel}.
This kernel is dependent on the relative distances between the observer and the lensed galaxies that are used to create the mass maps.
The source galaxy redshift distribution $n(z)$ used in this work is the non-tomographic distribution from ~\citep{fluri2019kids}.
The projected matter distribution is pixelised into images of size $128 \mbox{ px} \times 128 \mbox{ px}$, which corresponds to $5 \mbox{ deg} \times 5 \mbox{ deg}$ of the sky. 
Eventually, the resulting dataset consists of $57$ sets of $12'000$ sky convergence maps for a total of $684'000$ samples.
At training time, we randomly rotate and flip the input image to augment the dataset.

The dataset is split into a training and test set in the following way: $11$ cosmologies ($132'000$ samples) are selected for the test set, and the remaining $46$ cosmologies ($552'000$ samples) are assigned to the training set, as depicted in Figure~\ref{fig:grid_test_train}. 
This split is used to ensure that the model could interpolate to unseen cosmologies.
At evaluation time, we use the cosmologies from the test set to validate the interpolation ability of our network.
In the following sections, we show detailed summary statistics for the cosmologies marked with letters A,B,C, and D.
We make the dataset publicly available.\footnote{\url{https://zenodo.org/record/4646764}}

\section{Quantitative comparison metrics}
\label{sec:metrics}

We make a quantitative assessment of  the quality of the generated maps using both cosmological summary statistics and similarity metrics used in computer vision.
We focus on the following statistics: 
\begin{enumerate}
\setlength{\itemsep}{0.5em}
\item the power spectral density $C_{\ell}$, which describes how strongly the maps are correlated as a function of pixel separation $\ell$,
\item the distribution of mass map pixels $N_{\rm{pixels}}$, compared using Wasserstein-1 distance and histograms,
\item the distribution of mass map peaks $N_{\rm{peaks}}$, which describes the distribution of values at the local maxima of the map, compared also compared using Wasserstein-1 distance and histograms,
\item the bispectrum $B_{\ell}$, which describes the three-point correlation of the folded triangles of different size, 
\item Minkowski functionals, which are morphological measures of the map, and consist of three functions: $V_0$, which describes the area of the islands after thresholding of the map at some density level,  $V_1$, their perimeter, and $V_2$, their Euler characteristic (their number count minus the number of holes),
\item the Pearson's correlation matrices $R_{\ell\ell^\prime}$ between the $C_{\ell}$ of maps at different cosmologies,
\item the Multi-Scale Structural Similarity Index (MS-SSIM) \cite{wang2003multiscale,odena2017conditional}, which is an image similarity measure commonly used in computer vision,
\item the Fr\'echet Distance between the output of a CNN regressor trained to predict $\Omega_m,\sigma_8$, similarly to the Fr\'echet Inception Distance calculated using the Google Inception v3 network \citep{Hausel2017gans}.
\end{enumerate}
The mass map histograms and the peak counts are simple statistics used to compare the maps and constrain cosmological models (see \cite{Gatti2020moments,Kacprzak2016peaks} for examples).
These metrics, however, ignore the spatial information in the maps.
The angular power spectrum $C_\ell$ or its real-space equivalent, the angular correlation function, is the most common statistic for constraining cosmology with LSS (see \citep{Kilbinger2015review} for review).
The 2-pt functions capture only the Gaussian part of the fluctuations.
The 3-pt correlation function, or the bispectrum, probes higher order information and has also been used for constraining cosmological models \cite{Takada2003three,Fu2014three}.
Similarly, the Minkowski functionals have also been used for cosmological measurements \cite{Petri2015minkowski} as an alternative statistic that extracts topological information from the maps.

The agreement between the pixel and peak values of N-body and GAN-generated images is quantified using the Wasserstein-1 distance $W_1(P,Q)$.
This distance corresponds to the {\it optimal transport} of probability mass to turn the distribution $P$ into $Q$.
As it is scale-dependent, we calculate it after normalizing the pixel values: we subtract the mean and divide by the standard deviation.
We use mean and standard deviation of all N-body generated images for a given cosmology, for both samples.
This way, the $W_1$ distance is easily interpretable: for a Gaussian with $\mu{=}0, \sigma{=}1$, a 1$\sigma$ shift of the mean corresponds to $W_1{=}1$, and scaling its variance by $\times 2$ lead to $W_1{\approx}0.8$.

For $B_{\ell}$, $C_{\ell}$, and $V_{0,1,2}$ we calculate the simple fractional difference between the original and generated samples, defined as $f_{x} = (x_{\rm{GAN}}/x_{\rm{N-body}})/x_{\rm{N-body}}$. 
We quantify the agreement between correlation matrices by comparing their Frobenious norms $\norm{\cdot}_F$.
For a N-dimensional, diagonal covariance matrix with elements $\sigma_i^2$, the Frobenious norm scales linearly with $\sigma/\sqrt{N}$.
This way, it can be interpreted as a linear proxy for information content.
We define the fractional difference between the Frobenious norm of GAN and N-body correlation matrices as:
\begin{equation}
\label{eqn:fro_norm_diff}
f_{\rm{\bf R}}= \frac{\norm{\rm{\bf R}^{GAN}}_F - \norm{\rm{\bf R}^{N-body}}_F}{ \norm{\rm{\bf R}^{N-body}}_F }.
\end{equation}

The Multi-Scale Structural Similarity Index (MS-SSIM) is useful in order to detect the problem commonly known as mode collapse, where the generator produces only a small subset of the training data distribution.
Detecting this undesirable behaviour is non-trivial as summary statistics can still agree during mode collapse. 
Taking inspiration from \cite{odena2017conditional}, one solution is to leverage the MS-SSIM score from \cite{wang2003multiscale} to quantify this effect. 
This metric was first proposed for prediction of similarity in human perception of images. 
Taking two images as inputs, it returns a value between $0$ and $1$, where $1$ means ``identical'' and $0$ means ``completely different''.
As the mass maps are stochastic and only similar in a statistical way, we are not interested in the similarity between a pair of specific images, but in the average similarity of a large set of images. 
We calculate the significance of the difference in the SSIM measures in the following way:
\begin{equation}
\label{eqn:sssim_significance}
s_{\rm{SSIM}} = \frac{\langle \rm{SSIM}^{GAN} \rangle- \langle \rm{SSIM}^{N-body} \rangle} { (\sigma[\rm{SSIM}^{GAN}] + \sigma[\rm{SSIM}^{N-body}])/2},
\end{equation} 
where $\langle \rm{SSIM} \rangle$ is the mean score, and $\sigma[\rm{SSIM}]$ is the standard deviation. 
Large differences in the SSIM score indicate a significant difference in the samples generated by the GAN, thus pointing out to potential problems with the quality of the generated samples.
On the other hand, a small difference will be an indicator that the generative model preserve the data statistics.

Finally, we calculate an adaptation of the Fr\'echet Inception Distance (FID)~\cite{heusel2017gans}  between N-body and GAN -generated images.
The Inception Score (IS)~\cite{salimans2016improved} and FID have become standard measures for GANs.
The idea consists to compare statistics of the output of the Google Inception-v3 network~\cite{szegedy2016rethinking} for the ImageNet dataset~\cite{deng2009imagenet}. 
This has proven to be well correlated with human score. 
As the reference Inception network used for the FID was trained with the ImageNet dataset, its output statistics are meaningless for cosmological mass maps. To solve this challenge, we create our own reference network that is well suited for cosmological mass maps.
This network is a CNN trained to perform a regression task and predict the true $\sigma_8, \Omega_m$ parameters, similarly to \cite{schmelzle2017cosmological,fluri2018cosmological, gupta2018non}. Its parameters and detailed explanations of its construction can be found in Table~\ref{tab:regressor_model} and in Appendix B.
The adapted FID score is obtained by comparing the regressor outputs for the N-body and GAN images. As regressor is composed of 7 layers, this comparison depends on high order moments.
Naturally, we expect that a well working conditional GAN should generate samples with similar output distribution to the one of the real samples. 
To estimate the distance between the two statistics distributions, we first approximate the network predictions with a normal distributions $\mu_r, \Sigma_r$ and $\mu_g, \Sigma_g$, for the N-body and GAN -generated input, respectively.
The FID is then calculated as:
\begin{equation}
\label{eqn:fid_calculation}
\text{FID} = \|\mu_r - \mu_g\|^2 +  \| \Sigma_r^{1/2} - \Sigma_g^{1/2}\|^2.
\end{equation}
Note that this formula also correspond to the Wasserstein-1 distance between the two Gaussian distributions \cite{Dowson1982frechet}.
Eventually, before calculating FID, we normalise the network outputs for each true cosmology: we subtract the mean and divide by the standard deviation of the N-body sample.
For the ease of interpretation, we report the square root of FID.
This way, a $1\sigma$ difference in the mean CNN predictions will correspond to FID$^{1/2}$=1.
Similarly, a change of $1\sigma$ in the covariance matrix also leads to FID$^{1/2}$=1.

\section{Results}
\label{sec:kids_conditional_results}

\begin{figure*}[b]
  \centering
  \includegraphics[width=0.98\textwidth]{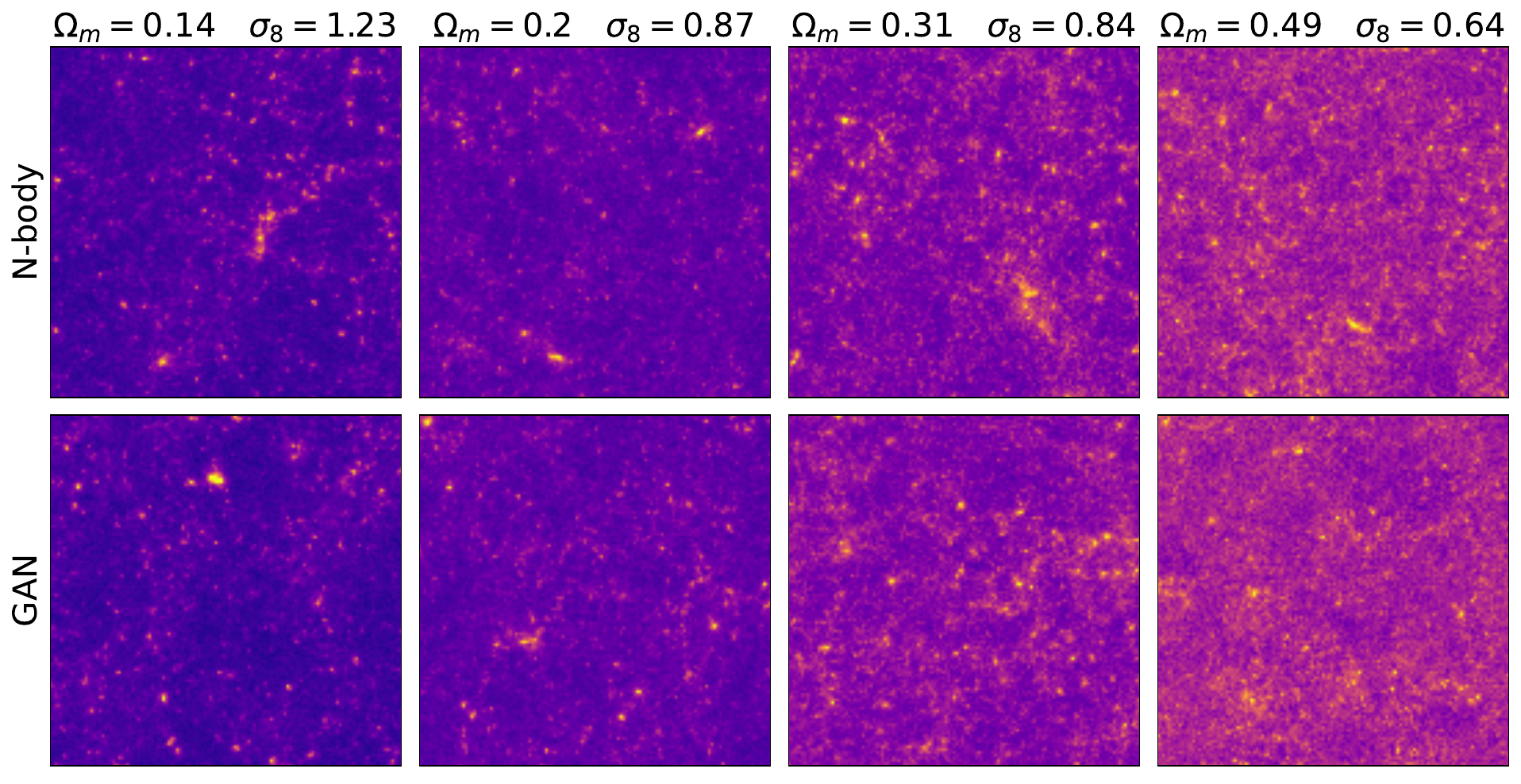}
  \caption{The original N-body images and GAN-generated images for four cosmological parameter sets.}
  \label{fig:kids_conditional_img}
\end{figure*}

We trained the GAN model described in Section~\ref{sec:conditional_gans} and Appendix A.
We used \textsc{RMSProp} as an optimizer with an initial learning rate of $10^{-5}$ and a batch size of $64$. 
The discriminator was updated $5$ times more than the generator.
The gradient penalty was set to $10$ and the negative slope of the LeakyRelu $\alpha = 0.2$.
It took a week to train the model for $40$ epochs on a GeForce GTX 1080 GPU. 
Similar to \cite{mirza2014conditional, reed2016generative, miyato2018cgans, odena2017conditional}, we use batches composed of samples from different parameter sets.
Note that the batches were composed of samples from different cosmologies from the \emph{training} set.
The summary statistics are computed using $5000$ real and fake samples for every pair of parameters of the \emph{test} set. 
The peaks are extracted by searching for all pixels greater than their $5 \times 5 $ patch neighborhood, i.e. their $24$ neighbours.
Then, the histogram of the extracted peaks values is computed.
We rely on \textsc{LensTools}~\cite{lenstools} to compute the power spectra, bispectra and the Minkowski functionals.
For the bispectrum, we use the {\em folded} configuration, with the ratio between one of the triangle sides and the base is set to the default value~of~0.5. 
The SSIM is computed using the \textsc{Scikit-Image} packge\footnote{\url{scikit-image.org}} \citep{van2014scikit}.
We make our code is publicly available\footnote{\url{https://renkulab.io/gitlab/nathanael.perraudin/darkmattergan}}.

\begin{figure*}
  \centering
  \includegraphics[clip, trim=0cm 0cm 0cm 0cm, width=0.6\linewidth]{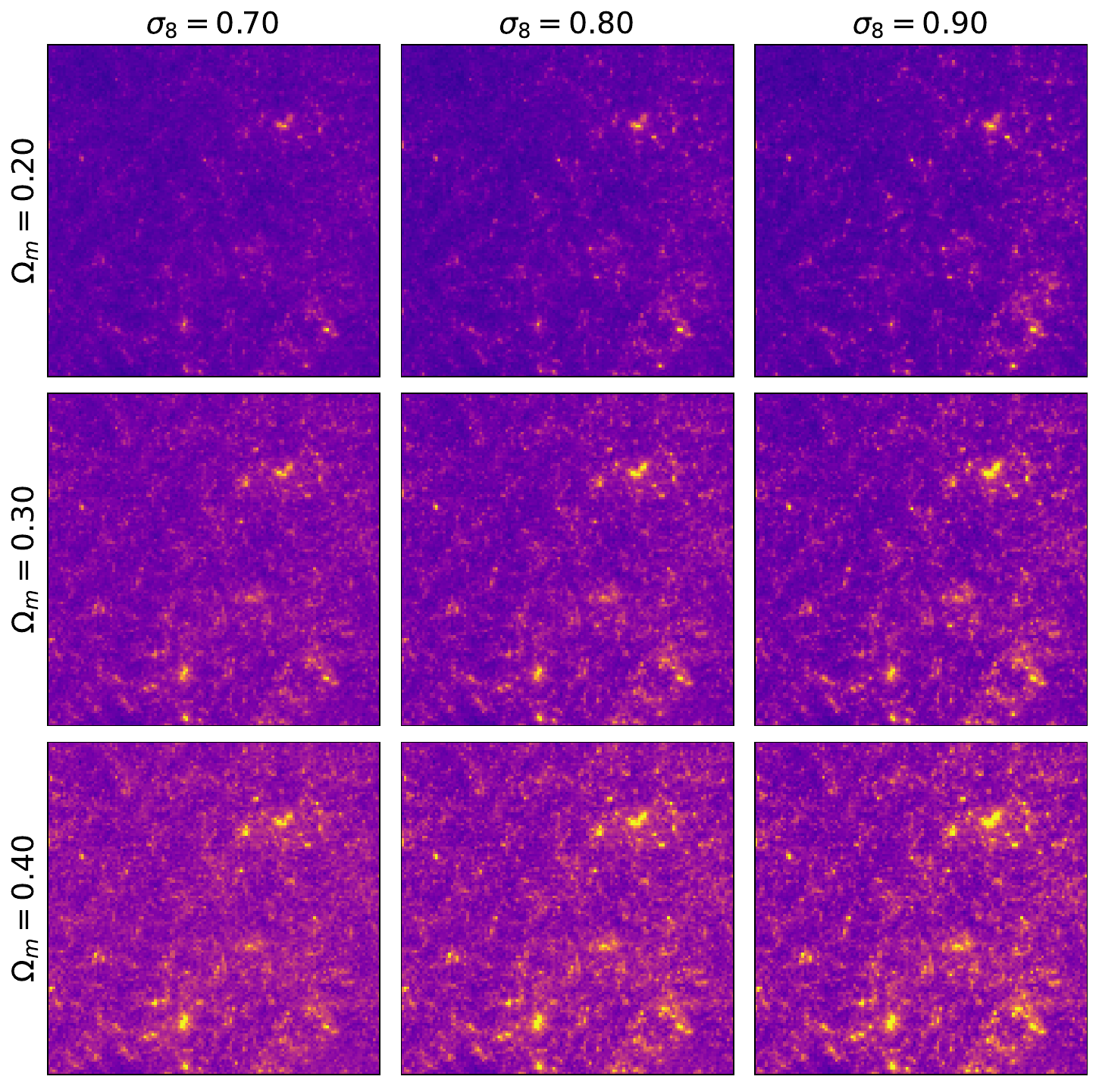}
  \caption{Images generated with the same random seed but with different input cosmology parameters $\Omega_m$ and $\sigma_8$.}
  \label{fig:kids_conditional_morphing}
\end{figure*}

Figure~\ref{fig:kids_conditional_img} shows images generated by the conditional GAN and as well as original ones, for several values of $\Omega_m$ and $\sigma_8$ parameters.
They are visually indistinguishable. 
Furthermore, the image structure evolves similarly with respect of the cosmological parameters change.
As predicted by the theory, increasing $\Omega_m$ results in convergence maps with additional mass and increasing $\sigma_8$ in images with higher variance in pixel intensities. 
In Figure~\ref{fig:kids_conditional_morphing} the same latent variable $z$ is used to generate different cosmologies. 
The smooth transition from low to high mass density hints that the latent variable control the overall mass distribution and the conditioning parameter its two cosmological properties $\sigma_8,\Omega_m$.

\begin{figure}[!htb]
  \centering
  \includegraphics[clip, trim=5cm 0cm 5cm 0cm, width=1\textwidth]{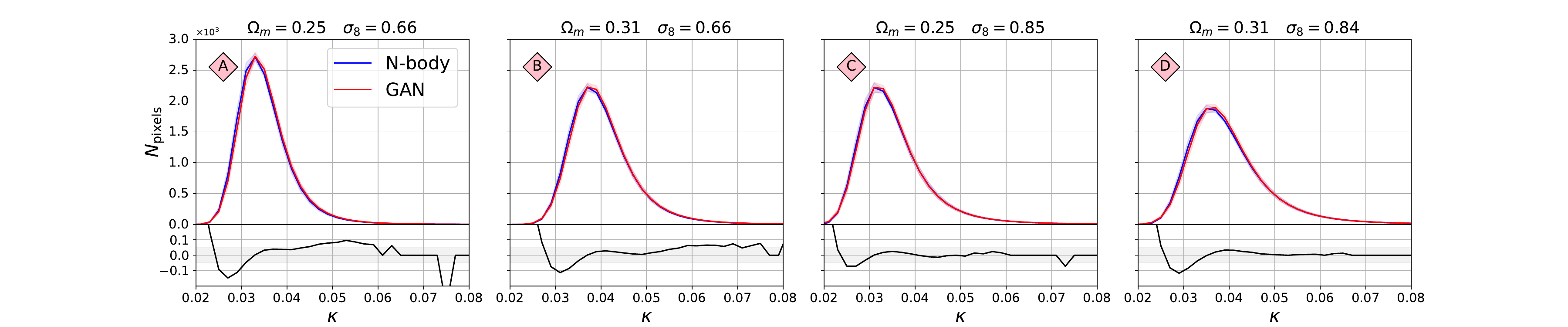}
  \includegraphics[clip, trim=5cm 0cm 5cm 0cm, width=1\textwidth]{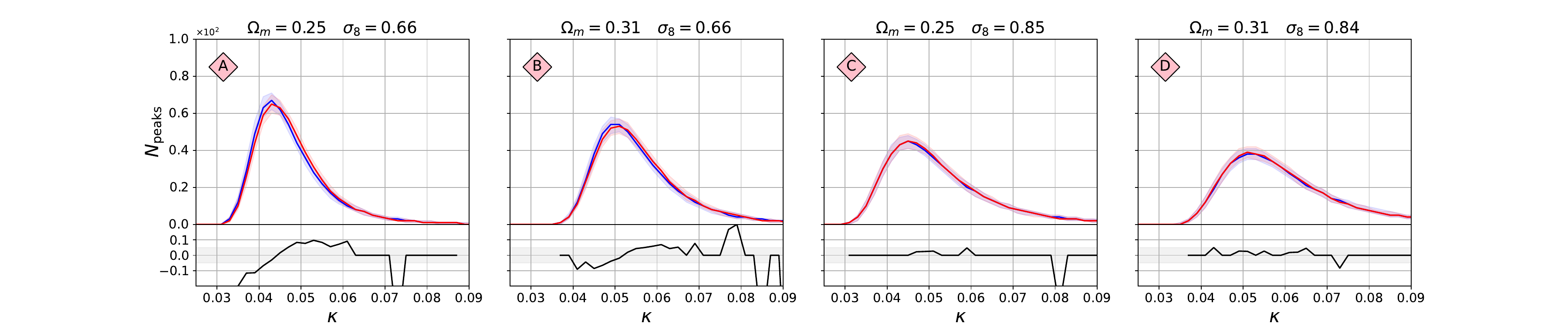}
    \caption{\label{fig:kids_conditional_stats_inter_a}  
    Comparison of histogram of pixel values (top) and peaks (bottom) between the original N-body and the GAN-generated maps. 
    Models A,B,C,D correspond to the ones marked in Figure~\ref{fig:grid_test_train}.
    The x-axis value $\kappa$ is the map pixel intensity.
    The solid line is the median histogram from 5000 randomly selected maps. 
    The bands correspond to $32\%$ and $68\%$ confidence limits of the ensemble of histograms.
    The lower panels show the fractional difference between the median histograms.
    }
  \includegraphics[clip, trim=5cm 0cm 5cm 0cm, width=1\textwidth]{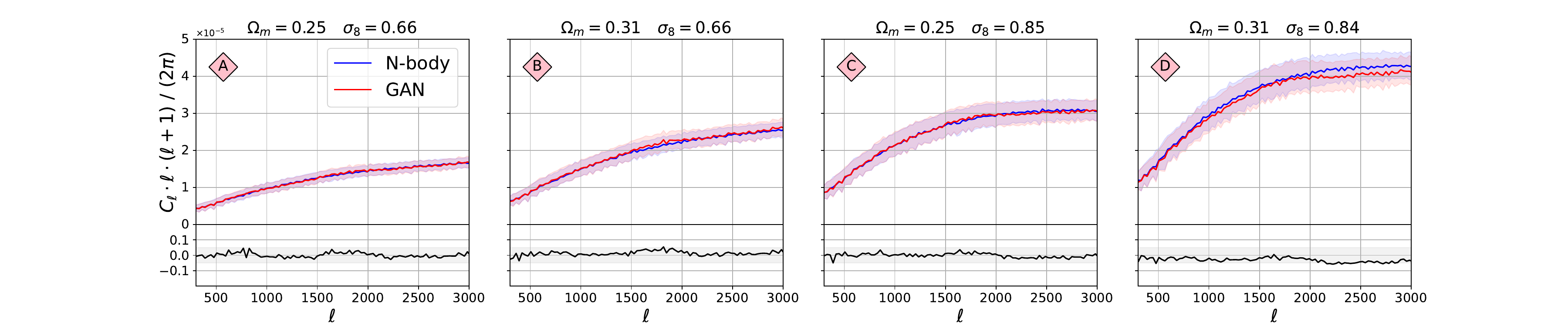}
  \includegraphics[clip, trim=5cm 0cm 5cm 0cm, width=1\textwidth]{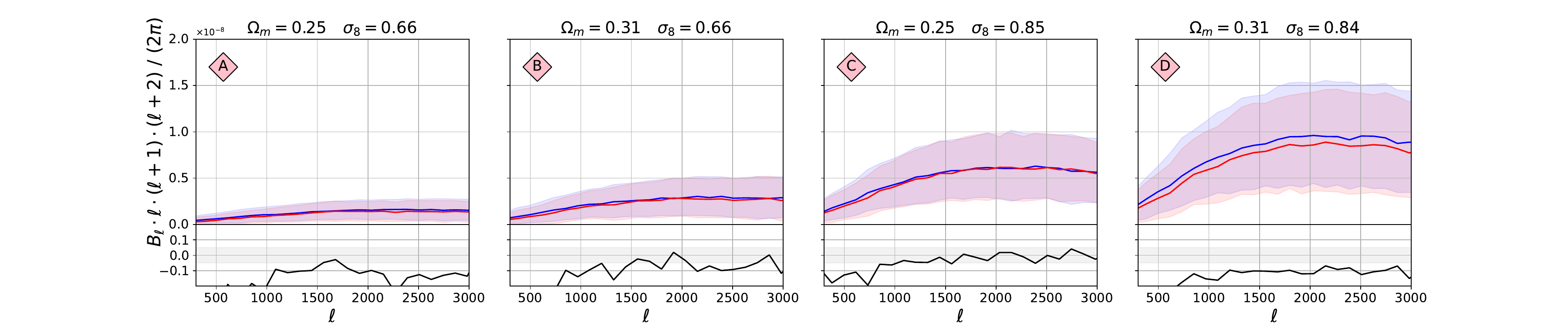}
      \caption{\label{fig:kids_conditional_stats_inter_b}  
      Comparison of the 2-pt and 3-pt functions between the original N-body maps and GAN-generated maps.
      The structure of this figure is the same as for Figure~\ref{fig:kids_conditional_stats_inter_a}.
      }
\end{figure}

Figure \ref{fig:kids_conditional_stats_inter_a} shows the histograms of pixels (top) and peaks (bottom) of the original maps simulated using  N-body simulations (blue), and their GAN-generated equivalents (red), for the four models A,B,C,D shown in Figure~\ref{fig:grid_test_train}.
The peak counts were selected as maxima of the surrounding 24 neighbours.
The solid line corresponds to the median of the histograms from 5000 realisations, and the bands to $32\%$ and $68\%$ percentiles.
The bottom part of each panel shows the fractional difference between the statistics, defined as $f_{x} = (x_{\rm{GAN}}/x_{\rm{N-body}})/x_{\rm{N-body}}$.
The normlised Wasserstein-1 distance of the pixel values distribution (see Section~\ref{sec:metrics}) is:
$
W_{1}^{\rm{pixel}}=
0.04,
0.02,
0.02,
0.02,
$
for models A,B,C, and D, respectively.
That indicates that the histograms differ on the level of $<5\%$.
Similarly, the Wasserstein-1 distances of the peak value distribution for models A,B,C, and D is:
$
W_{1}^{\rm{peak}} = 
0.04,
0.03,
0.01,
0.02.
$
The agreement here is also very good, on $<5\%$ level.

\begin{figure*}
\begin{center}
  \centering
  \includegraphics[clip, trim=5cm 0cm 5cm 0cm, width=1\textwidth]{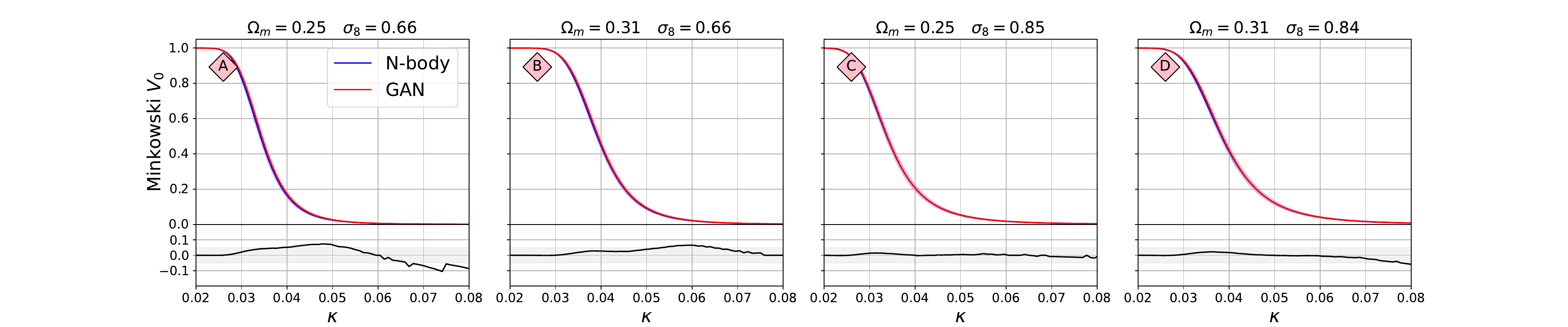}
  \includegraphics[clip, trim=5cm 0cm 5cm 0cm, width=1\textwidth]{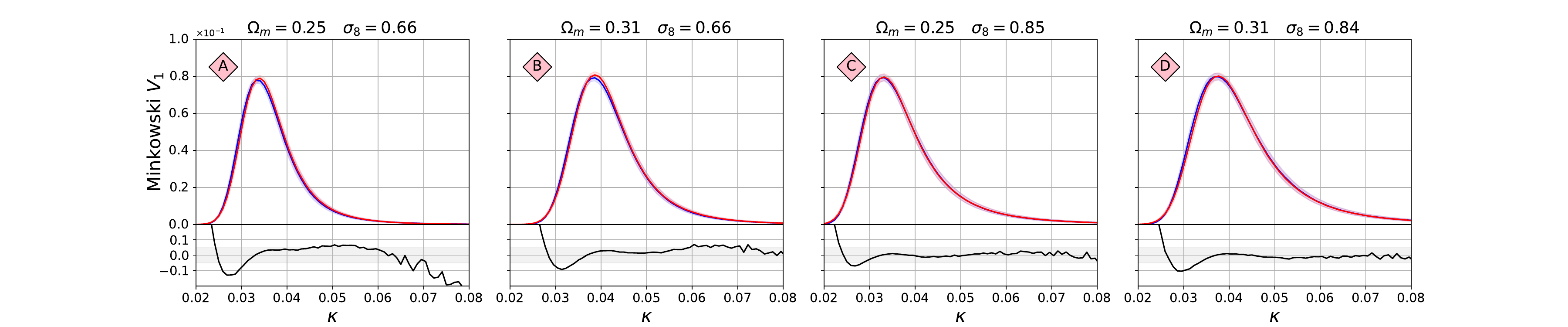}
  \includegraphics[clip, trim=5cm 0cm 5cm 0cm, width=1\textwidth]{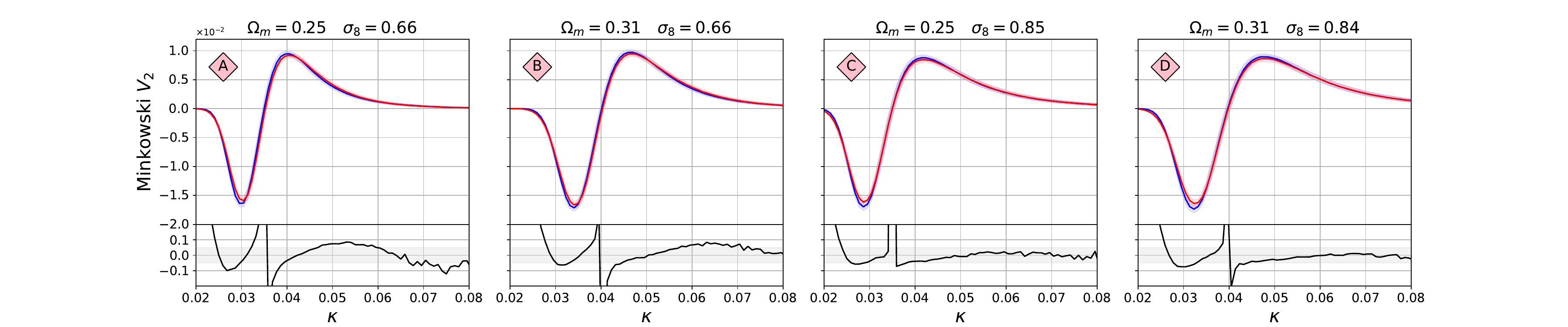}
      \caption{
      Comparison of the Minkowski functional between the original N-body maps and GAN-generated maps.
      The value of threshold $\kappa$, above which the functional value is calculated, is shown on the x-axis.
      The functional $V_0$ corresponds to the area of the emerging ``islands'', $V_1$ to their circumference, and $V_2$ to their Euler characteristic (their number count minus the number of holes in them). 
      The structure of this figure is the same as for Figure~\ref{fig:kids_conditional_stats_inter_a}.
      }
    \label{fig:kids_conditional_stats_inter_c}  
\end{center}
\end{figure*}

\begin{figure*}
  \centering
  \includegraphics[width=0.98\textwidth]{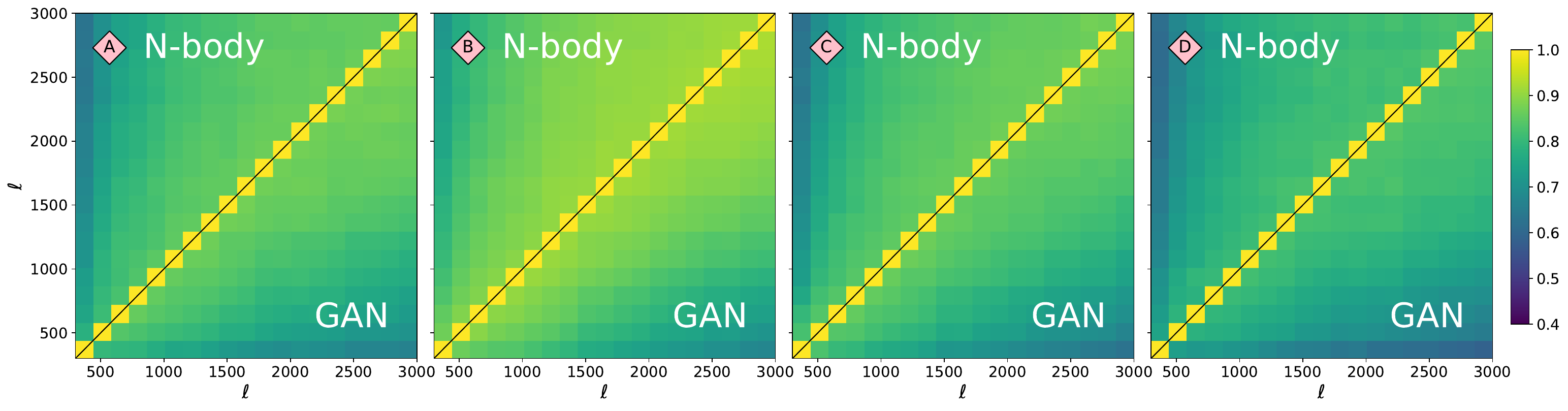}
  \caption{
  Pearson's correlation matrices for the four models highlighted in Figure~\ref{fig:grid_test_train}.
  The upper triangular corresponds to the original N-body power spectra, while the lower triangular to the power spectra of GAN-generated images.
  The fractional difference of the Frobenious norms (Equation~\ref{eqn:fro_norm_diff}) of these matrices is 
$
f_{\rm{\bf R}}=
0.02,
0.14,
0.06,
0.06
$, for models A,B,C, and D, respectively.
  }
  \label{fig:correlation_matrices}
\end{figure*}
The 2-pt and 3-pt statistics are shown in Figure~\ref{fig:kids_conditional_stats_inter_b}.
The power spectra $C_{\ell}$ overlap almost perfectly for all the cosmologies lying inside the parameter grid used for training.
Again, the agreement is better than $5\%$.
The agreement for the bispectrum $B_{\ell}$ is good for models B and C, but worse for A and B; the GAN model seems to underestimate the strength of the 3-pt correlations for these models, which differ by $\approx 20\%$.
We note that the 3-pt signal is very weak and has a large variation, which may be difficult to model for the GANs.

The Minkowski functionals are presented in Figure~\ref{fig:kids_conditional_stats_inter_c}. They were calculated using \textsc{Lenstools} \cite{lenstools}. The functional $V_0$ (first line) corresponds to the area of the emerging ``islands'', $V_1$ (second line) to their circumference, and $V_2$ (third line) to their Euler characteristic (their number count minus the number of holes in them). The value of threshold $\kappa$, above which the functional values, i.e. the ``islands'' are calculated, is shown on the x-axis. 
Here the agreement is typically better than $10\%$, with some model D agreeing much better, to $\approx 2\%$.
The large differences in the fractional difference plots are due to instability close to value of $V=0$.
The confidence limits of the summary statistics shown in these figures overlap very well, which indicates that the variability of these statistics is also captured very well by the GAN system.

The Pearson's correlation matrices $\rm{\bf R}$ of the power spectra are shown in Figure~\ref{fig:correlation_matrices}.
Those correlations were created from a coarsely-binned $C_{\ell}$ in range $\ell \in [300,3000]$.
The upper and lower triangular parts of the matrix show the original N-body correlations and the GAN correlations, respectively.
We calculate the Frobenious norms of these matrices and compare their ratios using Equation~\ref{eqn:fro_norm_diff}.
For the models A,B,C,D this difference is: 
$
f_{\rm{\bf R}} =
0.02,  
0.14,  
0.06,  
0.06.      
$
This agreement is overall very good, with model B being slightly worse.
As the precision requirements for covariance matrices are not as strict as for the summary statistics, this level of agreement can be considered satisfactory for upcoming applications \cite{Taylor2013precision}.

We calculate the mean and standard deviation of MS-SSIM score between 5000 randomly selected images for each cosmology, both for GAN and original N-body maps.
We test if the mean SSIM score is consistent between the N-body and GAN data using Equation~\ref{eqn:sssim_significance}.
The SSIM difference significance for the four models A,B,C, and D are:
$
s_{\rm{SSIM}} =
 0.08,  
 0.23,  
-0.27,  
 0.42,      
$.
This indicates very good statistical agreement for these models.

\begin{figure*}
  \centering
  \includegraphics[width=0.5\textwidth]{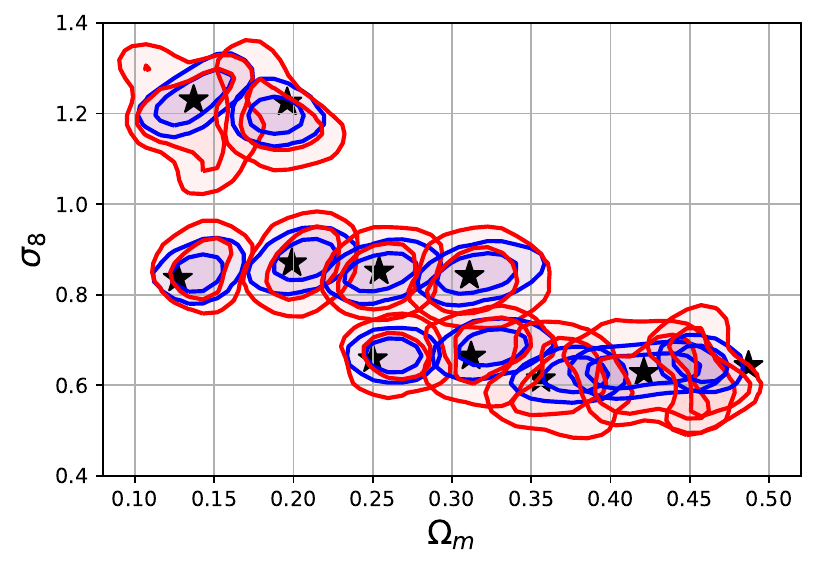}
  \caption{
  Predictions of a regressor CNN trained to predict the $\Omega_m, \sigma_8$ from input images.
  The details of this experiment are described in Section~\ref{sec:metrics} and the network architecture in Table~\ref{tab:regressor_model} in Appendix A.
  The contours encircle the 68\% and 95\% samples for the N-body maps (blue) and GAN-generated maps from the test set.
  The black stars show the true values of the test set parameters.
  }
  \label{fig:regression}
\end{figure*}

Figure~\ref{fig:regression} shows the prediction of a regressor CNN trained on the N-body images with true $\sigma_8, \Omega_m$ values.
For each category, we make the prediction with 500 randomly selected maps.
The shaded areas show the 68\% and 95\% probability contours for the N-body image input (blue) and the GAN image input (red).
The agreement is relatively good, but differences in the spread of these distributions is noticeable.
The Fr\'echet Distance (FID) computed using the reference cosmological CNN, as described in Section~\ref{sec:metrics}, is:
$
\text{FID}^{1/2}=
1.26,
1.44,
1.00,
1.19,
$
for models A,B,C, and D.
This indicates a slight difference according to this metric and agree with the distributions in Figure~\ref{fig:regression}.

\begin{figure*}
  \centering
%
  \includegraphics[clip, trim=0cm 0cm 0cm 0cm, width=1\linewidth]{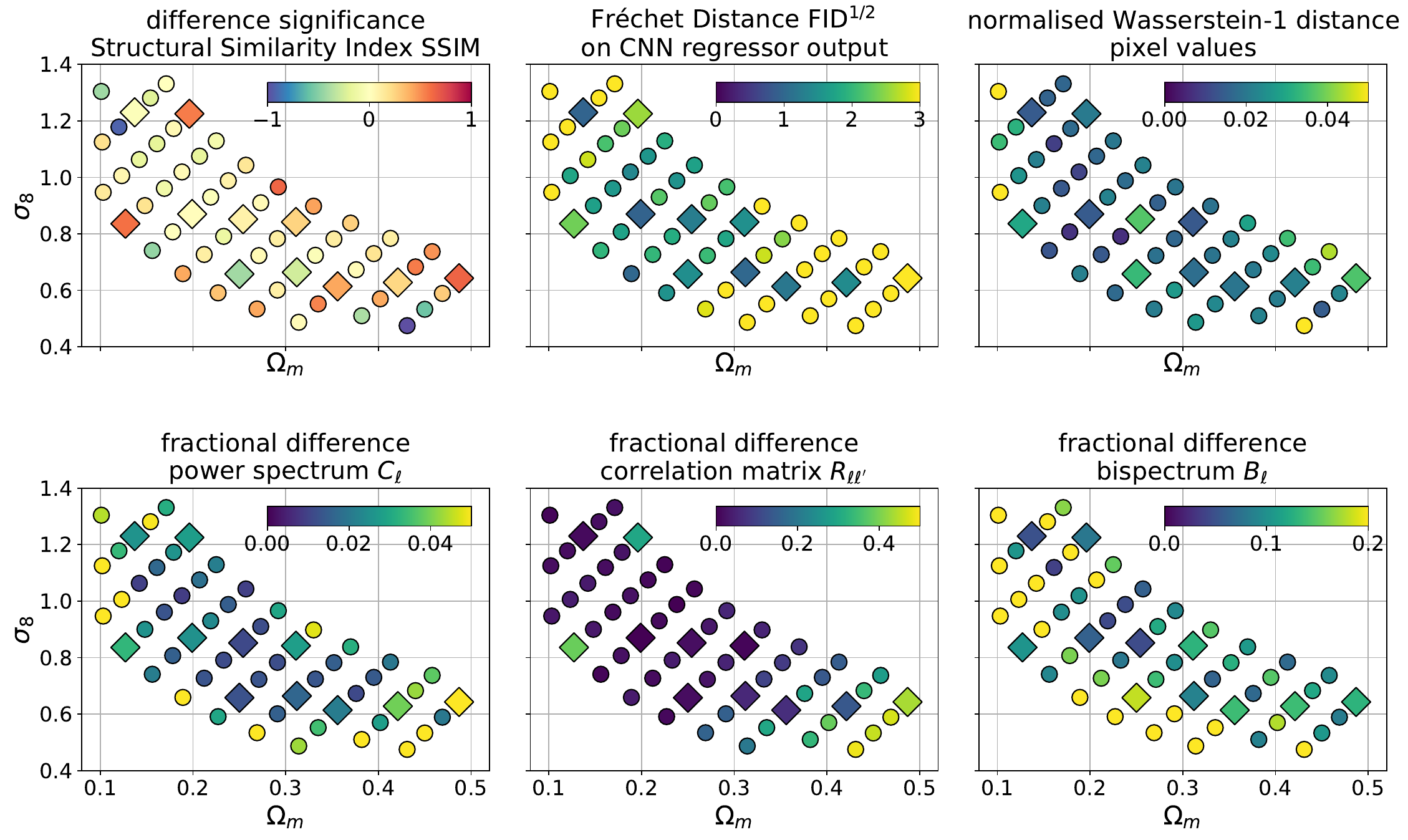}
  \caption{
  Differences between summary statistics of the original N-body and the GAN-generated images.
  The left panel shows the significance of the difference in Multi-Scale Structural Similarity Index (MS-SSIM), defined in Equation~\ref{eqn:sssim_significance}. 
  The upper middle panel presents the Fr\'echet Distance, computed using a regressor CNN and Equation~\ref{eqn:fid_calculation} (see Section~\ref{sec:metrics}).
  The upper right panel shows the normalised Wasserstein-1 distance in the pixel value distributions (see Section~\ref{sec:metrics}).
  The lower left panel shows the mean absolute fractional difference of the power spectra $C_\ell$.
  The lower middle panel shows the fractional difference in the Frobenius norms of correlation matrices, defined in Equation~\ref{eqn:fro_norm_diff}.
  The lower right panel present the mean absolute fractional difference of the bispectrum $B_{\ell}$.
  The circles and squares indicates parameters from the training and the test sets.
  }
  \label{fig:stats_difference_grid}
\end{figure*}

We compare the summary statistics as a function of cosmological parameters for both the training and the test set.
We used the training and test sets displayed in Figure~\ref{fig:grid_test_train}. 
Figure~\ref{fig:stats_difference_grid} shows the six quantities as a function of cosmological parameters: 
\begin{itemize}[leftmargin=0cm]
    \item[top left:] significance of the difference $s_{\rm{SSIM}}$ in Multi-Scale Structural Similarity Index (Eqn.~\ref{eqn:sssim_significance}),
    \item[top center:] Fr\'echet distance using a CNN regressor (Eqn.~\ref{eqn:fid_calculation}). Note that for a Gaussian distribution, a difference of FID$^{1/2}$=1 corresponds to either a $1\sigma$ shift in the mean or $1\sigma$ difference in standard deviation,
    \item[top right:] normalised Wasserstein-1 distance of the pixel value distribution. For a Gaussian distribution, a $1\sigma$ change in the mean corresponds to $W_1{=}1$, and an increase in standard deviation of $\times 2$ to $W_1{\approx}0.8$,
    \item[bottom left:] average fractional differences in the power spectrum $f_{C_{\ell}}$,
    \item[bottom center:] fractional differences in the power bispectrum $f_{B_{\ell}}$,
    \item[bottom left:] fractional difference in the Frobenious norm of the correlation matrices $f_{\rm{\bf R}}$ (Eqn.~\ref{eqn:fro_norm_diff}).
\end{itemize}

Overall, we notice that the agreement between the N-body simulations and GAN-generated maps is the best in the centre of the grid for both the training and test set.
The fact that the differences in neighbouring cosmologies are similar indicates that the GAN system can efficiently learn the latent interpolation of the maps.
The agreement worsens at the edges of the grid. 
We observe the biggest deterioration in the realism of the GAN model for the high $\Omega_m$ and low $\sigma_8$ parameters.
This is most prominent for the correlation matrix and the SSIM differences.
Conversely, the biggest difference for the bispectrum is present for low $\Omega_m$ and high $\sigma_8$.

\section{Conclusion}
\label{sec:conclusion}
We proposed a new conditional GAN model for continuous parameters where conditioning is done in the latent space. 
We demonstrated the ability of this model to generate sky convergence maps when conditioning on the cosmological parameters $\Omega_m$ and $\sigma_8$. 
Our model is able to produce samples that resemble samples from the test set with good statistical accuracy, which demonstrates its generalization abilities.
The agreement of the low order summary statistics (pixel and peak histograms and power spectrum) is very good, typically on the $<5\%$ level.
Higher order statistics (Minkowski functionals, bispectrum) agree well, but with larger differences, generally around $\approx 10\%$, and in some cases $\approx 20\%$.
The comparison of the Multi-Scale Structural Similarity Index (MS-SSIM) shows a good agreement in this metric, with the exception of the low $\sigma_8$ and high $\Omega_m$ edge of the grid.
Moreover, the GAN model is able to capture the variability in the conditioned dataset:
we observe that the scatter of the summary statistics computed from an ensemble is very similar between the original and generated images.
The investigation of the correlation matrices of the power spectra also shows a good agreement, with a quality deteriorating close to the edges of the grid, especially for low $\sigma_8$ and high $\Omega_m$.
This is not unexpected, as the training set contains less information near the edges of the grid.
More investigation is needed to more close inspect the behavior of the generative model in these areas.
As generative models are rapidly growing in popularity in machine learning, we anticipate to be able to solve these problems in the near future.

Our results offer good prospects for GAN-based conditional models to be used as emulators of cosmology-dependent mass maps.
As these models efficiently capture both the signal and its variability, the map-level emulators could potentially be used for cosmological analyses.
They can accurately predict the power spectrum and its covariance, which is often unattainable in standard cosmological analyses \citep{Eifler2009covariance}.
It can also be used for non-Gaussian analyses of lensing mass maps, such as, for example, in \cite{Zuercher2020forecast,Parroni2020minkowski}.
Further experiments will be needed, however, to bring the generative models to a level where they can be of practical use in a full, end-to-end cosmological analysis.

In this paper, we have demonstrated the ability of generative AI models to serve as emulators of cosmological mass maps for a given redshift distribution of source galaxies $n(z)$. 
Generative models have also been shown to work directly on the full or sliced 3D matter density distributions \citep{perraudin2019cosmological,Troester2019painting,VillaescuseNavarro2020camels}.
The three dimensional generation of cosmological fields proves to be particularly difficult.
As most of the survey experiments publish their lensing catalogues and their corresponding redshift distributions, the generation of projected maps, as shown in this work, could be of direct practical use.
Another challenge will be posed by the large sky area of the upcoming surveys and their spherical geometry.
Spherical convolutional neural networks architectures have been proposed \cite{perraudin2018deepsphere,Krachmalnicoff2019healpix,McEven2021scattering}.
These architectures are expected to be easy to implement with generative models, which offers good prospect for the development of spherical mass map emulators.

\section*{Acknowledgements}
This work was supported by a grant from the Swiss Data Science Center (SDCS) under project \textit{DLOC:  Deep Learning for Observational Cosmology} and grant number 200021\_169130 from the Swiss National Science Foundation (SNSF).
We thank Thomas Hofmann, Alexandre R\'{e}fr\'{e}gier, and Fernando Perez-Cruz for advice and helpful discussions.
We thank the Cosmology Research Group of ETHZ and particularly Janis Fluri for giving us access to the dataset.
Finally, we thank the two anonymous reviewers who provided extensive feedback that greatly improved the quality of this paper.

\appendix

\section*{A. Generative Adversarial Network Architecture}
\label{sec:appendix_network}

\begin{table*}
\centering
\begin{tabular}{cccc}
\hline
Layer & Operation & Activation & Dimension \\ \hline
\multicolumn{3}{l}{\textit{Generator}} & \multicolumn{1}{l}{} \\ \cline{1-3}
$\hat{z}$ & equation \ref{eq:gaussian_length} &  & $b \times 128$ \\
$g_0$ & linear & Relu & $b \times {\color{blue} 256}$ \\
$g_1$ & linear & Relu & $b \times {\color{blue}512}$ \\
$g_2$ & linear & Relu & $b \times {\color{blue} 32768}$ \\
$g_3$ & reshape &  & $b \times 8 \times 8 \times 512$ \\
$g_4$ & deconv ($k=3 \times 3$, $s=2$) & Relu & $b \times 16 \times 16 \times {\color{blue} 256}$ \\
$g_5$ & deconv ($k=5 \times 5$, $s=2$) & Relu & $b \times 32 \times 32 \times {\color{blue} 128}$ \\
$g_6$ & deconv ($k=5 \times 5$, $s=2$) & Relu & $b \times 64 \times 64 \times {\color{blue} 64}$ \\
$g_7$ & deconv ($k=5 \times 5$, $s=2$) & Relu & $b \times 128 \times 128 \times {\color{blue} 32}$ \\
$g_8$ & deconv ($k=7 \times 7$, $s=1$)  & Relu & $b \times 128 \times 128 \times {\color{blue} 1}$ \\
\multicolumn{1}{l}{\textit{Discriminator}} & \multicolumn{1}{l}{} & \multicolumn{1}{l}{} & \multicolumn{1}{l}{} \\ \cline{1-3}
$X$ &  &  & $b \times 128 \times 128$ \\
$d_0$ & conv ($k=7 \times 7$, $s=1$) & LeakyRelu & $b \times 128 \times 128 \times {\color{blue} 32}$ \\
$d_1$ & conv ($k=5 \times 5$, $s=2$) & LeakyRelu & $b \times 64 \times 64 \times {\color{blue} 64}$ \\
$d_2$ & conv ($k=5 \times 5$, $s=2$) & LeakyRelu & $b \times 32 \times 32 \times {\color{blue} 128}$ \\
$d_3$ & conv ($k=5 \times 5$, $s=2$) & LeakyRelu & $b \times 16 \times 16 \times {\color{blue} 256}$ \\
$d_4$ & conv ($k=3 \times 3$, $s=2$) & LeakyRelu & $b \times 8 \times 8 \times {\color{blue} 512}$ \\
$d_5$ & reshape + concatenate &  & $b \times 32770$ \\
$d_6$ & linear & LeakyRelu & $b \times {\color{blue} 512}$ \\
$d_7$ & linear & LeakyRelu & $b \times {\color{blue} 256}$ \\
$d_8$ & linear & LeakyRelu & $b \times {\color{blue} 128}$ \\
$d_9$ & linear & LeakyRelu & $b \times {\color{blue} 1}$\\
\end{tabular}\\
\caption{Conditional GAN architecture.  $d_5$ is a layer that reshapes the tensor to a vector and then concatenates the conditioning parameters to it. Here $b$ is the batch size, $k$ the convolutional kernel size and $s$ the stride. The number of filters (convolution layer) and the number of neurons (linear layers) is shown in blue. }
\label{tab:kids_gan_conditional}
\end{table*}

~

\textbf{Model.}
Table \ref{tab:kids_gan_conditional} summarizes the architecture of the GAN system, i.e. the generator and the discriminator.
From the latent variable $z$ and the cosmological parameters $\sigma_8$ and $\Omega_m$, the generator starts by computing $\hat{z}$ using \eqref{eq:gaussian_length}. 
As a second step, $\hat{z}$ is transformed with three linear layers. i.e. a Multi Layer Perceptron (MLP), that outputs a $32768$ tensor ($g_0$ to $g_2$). The data is then reshaped to $8 \times 8 \times 512$ ($h_3$) and further transformed with four deconvolutional layers with stride $2$ and kernel sizes of $3 \times 3$ or $5 \times 5$ ($g_4$ to $g_7$). The last generator layer consists of a deconvolution with stride $1$ and kernel size $7 \times 7$ and it is intended to generate fine-grained details ($g_8$).
The discriminator is symmetric to the generator with two exceptions. First the parameters $\sigma_8$ and $\Omega_m$ are concatenated in $d_5$ just before the first linear layer. Second, an extra linear layer is added at the end of the discriminator ($d_9$) in order to recover a single output. All layers are separated by a LeakyRelu activation function with the parameter $\alpha = 0.2$ \cite{maas2013rectifier}.

\textbf{Training.}
The cosmological dataset described in Section~\ref{sec:data_kids} is used to train the GAN, where the batches are composed of samples from different cosmologies.
We select a Wasserstein loss, with a gradient penalty of $10$ \cite{arjovsky2017wasserstein}.
We use RMSProp as an optimizer with an initial learning rate of $10^{-5}$, and a batch size of $64$. 
The discriminator is updated $5$ times more often than the generator.
The model is trained for $10^{-5}$  epochs on a GeForce GTX 1080 GPU, which takes around $170$ hours.

\section*{B. Regressor training}

\begin{table}[h]
\centering
\begin{tabular}{cccc}
\hline
Layer & Operation & Activation & Dimension \\ \hline
$X$ &  &  & $b \times 128 \times 128$ \\
$h_0$ & conv ($k=7 \times 7$, $s=1$) & LeakyRelu & $b \times 128 \times 128 \times  {\color{blue} 32}$ \\
$h_1$ & conv ($k=5 \times 5$, $s=2$) & LeakyRelu & $b \times 64 \times 64 \times  {\color{blue} 64}$ \\
$h_2$ & conv ($k=5 \times 5$, $s=2$) & LeakyRelu & $b \times 32 \times 32 \times  {\color{blue} 128}$ \\
$h_3$ & conv ($k=5 \times 5$, $s=2$) & LeakyRelu & $b \times 16 \times 16 \times  {\color{blue} 256}$ \\
$h_4$ & conv ($k=3 \times 3$, $s=2$) & LeakyRelu & $b \times 8 \times 8 \times  {\color{blue} 512}$ \\
$h_5$ & reshape &  & $b \times 32768$ \\
$h_6$ & linear & LeakyRelu & $b \times  {\color{blue} 512}$ \\
$h_7$ & linear & LeakyRelu & $b \times  {\color{blue} 256}$ \\
$h_8$ & linear & LeakyRelu & $b \times  {\color{blue} 128}$ \\
$h_9$ & linear & linear & $b \times {\color{blue} 2}$\\
\end{tabular}\\
\caption{Architecture of the regressor. Here $b$ is the batch size, $k$ the convolutional kernel size and $s$ the stride. The number of filters (convolution layer) and the number of neurons (linear layers) is shown in blue.  The LeakyRelu activation uses the parameter $\alpha = 0.2$.}
\label{tab:regressor_model}
\end{table}

Given real and generated images, the general idea of the Frechet Inception Distance (FID) is to compute the distance between some of their complex statistics. For natural images, these statistics are given by the last layer, i.e. the logits, of a pre-trained Inception-V3 network~\cite{szegedy2016rethinking}. 
As these statistics are meaningless for our cosmological data, we build new ones using a carefully designed regressor. Given an image, the regressor is trained to predict the two parameters $\Omega_m$ and $\sigma_8$. We provide the regressor weights with the code to make our FID metric reusable.

\textbf{Data.} Naturally, we use the training dataset described in Section \ref{sec:data_kids}, i.e. $46$ different cosmologies composed by $12000$ images each. 
This training dataset is further randomly split into a regressor training set ($80 \%$) and a restressor test set ($20 \%$).

\textbf{Model.} The architecture of the regressor is described in Table \ref{tab:regressor_model}. It shares the same structure as the GAN discriminator.
It consists of a four convolutional layers followed by three linear layers with leaky relu non-linearity. The last layer is a linear layer with two outputs and it is responsible for producing the predicted parameters. We select the LeakyRelu activation functions for better gradient propagation. 

\textbf{Training.}
We use the mean squared error between the predicted and true parameters as a loss function.
The  model was trained for $20$ epochs using an Adam~\cite{kingma2014adam} optimizer with an initial learning rate of $3 \cdot 10^{-5}$, $\beta_1 = 0.9$, $\beta_2=0.999$ and $\epsilon = 10^{-8}$ and a batch size of $64$.
The mean squared error evaluated on the test set corresponds to $8.93 e^{-5}$, which is low enough for the purpose of computing the FID.

\clearpage

\bibliographystyle{frontiersinHLTHFPHY} 
\bibliography{bibliography}

\end{document}